\documentclass[prb,twocolumn]{revtex4}
\usepackage{graphicx}
\begin{document}
\title{Mobility edge scaling at semiclassical and dissipative Hall
transitions}
\author{Joel E. Moore}
\affiliation{Department of Physics,
University of California, Berkeley, CA 94720}
\date{\today}
\begin{abstract}
The universal anomalous diffusion scaling is obtained for the
semiclassical quantum Hall transition, which has been
argued to describe samples with dissipation or correlated impurities.
The results explain a discrepancy between existing numerics and the
expected scaling $\eta = 2 x_1 = \frac{1}{2}$, which is violated
because of a cancellation in the scaling function.  The crossover
with increasing observation time from semiclassical to quantum scaling
is shown to explain a recent experiment which finds different scaling laws
depending on how the localization length is extracted.
\end{abstract}
\maketitle

The electronic eigenstates of disordered systems at critical points
such as the quantum Hall plateau transition~\cite{chalker} or
metal-insulator transition~\cite{mirlin} are believed to have fractal
structure described by universal scaling laws, but analytic results
for such scaling laws are quite scarce.  This paper finds the exact
anomalous diffusion scaling in a standard semiclassical model for the
quantum Hall transition, extending previous numerical
studies~\cite{evers}.  Several
authors~\cite{shimshoni,kapitulnik,avishai} have argued that this
semiclassical limit is relevant to various experimental situations, and it
also shares some features with the
spin quantum Hall transition~\cite{senthil,gruzberg} in disordered
superconductors.  Here the semiclassical model is compared to recent
experiments~\cite{hohls} which appear to show both ``quantum''
and "classical" localization scaling laws in the same sample, depending on how the localization length is extracted.

The semiclassical model has recently appeared in studies of how
dissipative effects can modify the quantum Hall
transition~\cite{shimshoni,kapitulnik}.  Many experiments observe
localization length scaling more consistent with the classical value
$\nu = \frac{4}{3}$ than the expected quantum value $\nu \approx 2.35
\pm 0.05$~\cite{wei,huckestein}.
By suppressing tunneling and
introducing a finite dephasing length, dissipation increases the range
of chemical potentials where a semiclassical description applies, but
causes simple diffusion over a nonzero range around the critical
chemical potential.  The main experimental results
can be understood in a simple picture incorporating dissipation, without requiring a
new dissipation-dominated critical point other than diffusion.

The starting point of the analysis is a standard lattice model for classical motion on
percolation hulls or random level surfaces; quantum mechanics only
enters this model in properties like the density of states and
conductivity~\cite{evers,gurarie} which count the number of
trajectories.  The random level surface problem was first introduced
in the quantum Hall context as a useful but quantitatively incorrect
model for the integer transition~\cite{trugman,isichenko}.  Our
approach uses a mapping to a class of lattice polymers~\cite{moore} to
find time-dependent correlations in the semiclassical random level
surface problem and resolve a disagreement between existing numerics
and analytical work.

The results improve upon Monte Carlo calculations of
Evers~\cite{evers} on the random level surface problem, which
correctly found a deviation from simple scaling but did not reach the
extremely long paths ($\geq 10^4$ disorder correlation lengths) where
the true asymptotic scaling sets in.  There is a universal anomalous
diffusion exponent $\eta = \frac{1}{4}$ which characterizes the
non-Gaussian correlation of critical states.  This result differs from
the value $\eta = \frac{1}{2}$ predicted by a simple scaling argument
because the leading scaling term vanishes.  The restricted
``open-walk'' version of the problem, which is more convenient for
numerics~\cite{evers}, is shown analytically to have $\eta =
\frac{1}{4}$ with divergent logarithmic corrections resulting from a
short-distance singularity in the associated polymer ensemble.  Monte
Carlo simulations are used to verify some of the predicted scaling
behavior.  Aside from their direct relevance to quantum Hall
transitions, the results suggest possible features of other
disordered electronic transitions: cancellations in scaling functions,
logarithmic corrections, and slow decay of finite-size effects.

The existence of anomalous scaling (nonzero $\eta$) for the ordinary
integer transition was shown by Chalker and Daniell~\cite{chalker}: in
the scaling regime $q,\omega \ll 1$, the spectral function has two
distinct universal limits:
\begin{eqnarray}
S({\bf q}; E_c, \omega) \sim \cases{q^2 \omega^{-2}&if $q^2 \ll
\omega$\cr \omega^{-\eta/2}q^{\eta-2}&if $q^2 \gg \omega$}.
\end{eqnarray}
This form satisfies the scaling law $S(q; E_c,\omega) =
\omega^{-1} f(q^2/\omega)$ which follows from a homogeneity
assumption.  $S(q; E_c,\omega)$ is
defined as the Fourier transform of
\begin{eqnarray}
S({\bf r}; E_c,\omega) &=& \Big\langle
\sum_{i,j} \delta(E_i + \frac{\omega}{2}
- E_c) \delta(E_j - \frac{\omega}{2} - E_c) \cr
&&\times \psi_i(0) \psi_i^*({\bf r}) \psi_j({\bf r}) \psi_j^*(0)\Big\rangle.
\label{sdef}
\end{eqnarray}
We now define the density-density correlation $\langle
\rho(0,0) \rho(r,t) \rangle$ at $E=E_c$ in the random level surface
model, with scaling analogous to (\ref{sdef}).

Consider a classical charged particle moving in the ${\bf x}$-${\bf y}$
plane in a magnetic field $B {\bf \hat z}$ and random potential
$V(x)$.  For a smoothly varying potential, the particle velocity
averaged over the fast cyclotron motion is ${\bf v} = ({\bf E}
\times {\bf B})/B^2$.  The particle velocity is perpendicular
to the potential gradient, and the particle moves along constant energy
surfaces of the random potential.  For a uniformly
distributed potential $V(x) \in [-1,1]$, the typical size of level
surfaces at energy $E$ diverges as $E \rightarrow 0$.  The connection
to percolation comes about because a level surface at energy $E$
separates regions with $V>E$ from those with $V<E$.  The level
surfaces are closed non-self-intersecting loops, whose statistical
properties are exactly the same as percolation hulls~\cite{trugman},
or self-interacting ring polymers at the critical
$\theta$-point~\cite{duplantier}.

\begin{figure}
\includegraphics[width=2.5in]{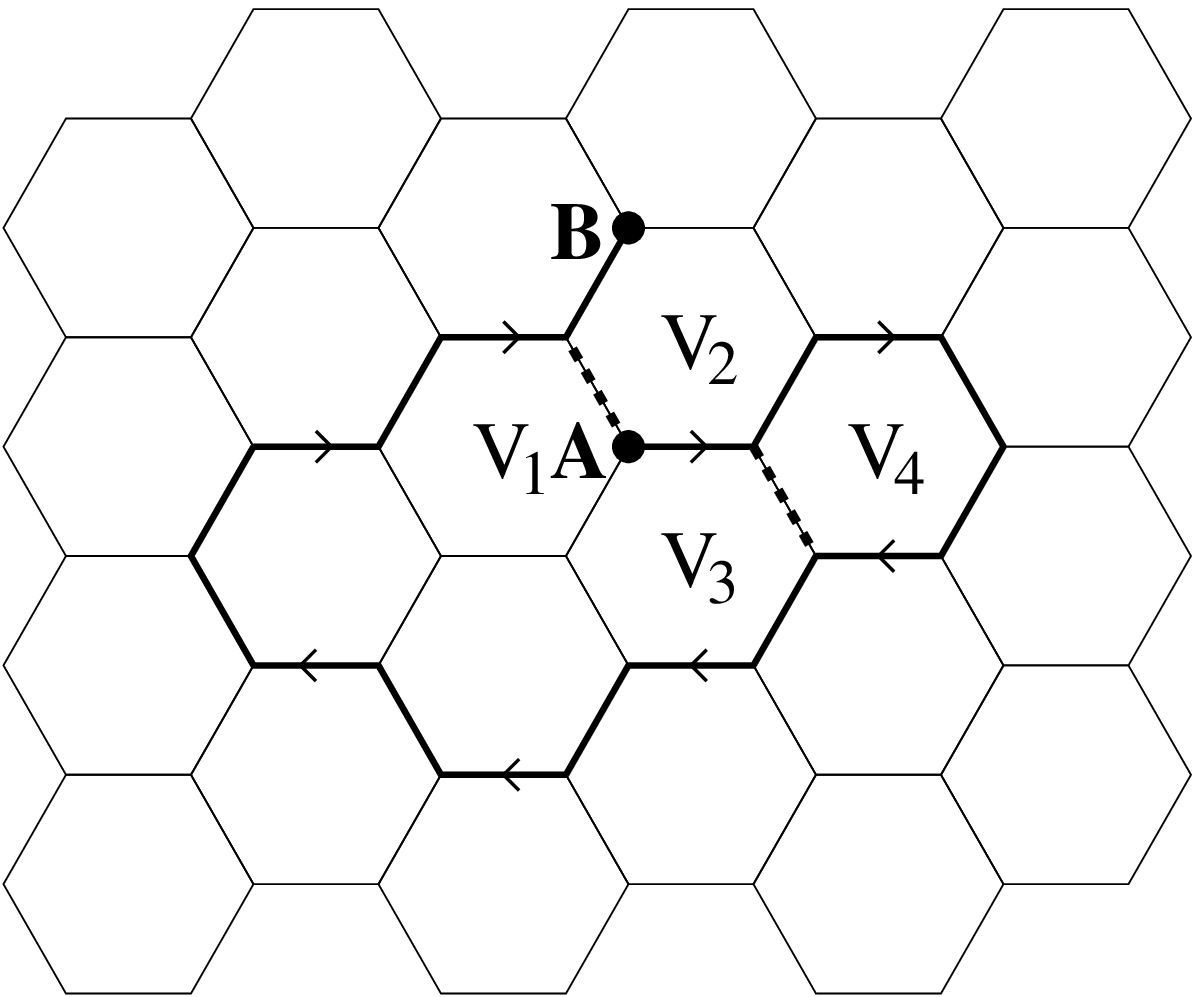}
\caption{The dotted edges are self-contacts:
the edge between $V_3$ and $V_4$ is an antiparallel self-contact,
while that between $V_1$ and $V_2$ is a parallel self-contact.
This path is not allowed classically since the walk passes
$V_2$ both on the right and on the left.}
\label{figone}
\end{figure}

The results from percolation and polymers used to study this problem
require discretizing the motion so that the particle moves on a
regular lattice.  From numerics~\cite{evers,gurarie} it is known that
the particle has nonzero mean velocity at the critical energy, so time
can be discretized as well: the particle takes one step on the lattice
per time unit.  The resulting model on the hexagonal
lattice~\cite{moore,gurarie} is depicted in Fig.~\ref{figone}.  There
is an independent random potential $V_i$ on each face of the
lattice.  A particle of energy $E$ moves so that the potential of
faces to its left (right) is always greater (less) than $E$.

Averaging over disorder, the probability to be at $b$ after $N$ steps
starting from $a$ can be written as a weighted sum over self-avoiding
walks (SAWs) or loops (SAPs)~\cite{moore}. The result is ($H$ is the
number of different hexagons visited by the SAW or SAP)
\begin{eqnarray}
f(r,N) &\propto& \sum_{{\scriptstyle{\rm SAPs\,}i{\rm \,through\,}
a {\rm\,and\,} b,\atop
\scriptstyle l = {\rm \,length\,of\,SAP,}} \atop
\scriptstyle q = {\rm \,steps\,from\,} a {\rm \,to\,} b}
\delta_{N {\rm\,mod\,} l, q} 2^{-H} \cr
&& + \sum_{{\scriptstyle{\rm SAWs\,}j{\rm \,of\,length\,}N \atop
\scriptstyle {\rm from\,} a {\rm\,to\,} b} \atop
\scriptstyle{\rm no\,}\parallel{\rm \,self-contacts}}
{2^{-H}}.
\label{ensemble}
\end{eqnarray}
Here each SAP should be summed twice, once with distance $q$ and once
with distance $l - q$.  The classical analogue of the spectral
function $S(q,\omega)$ is the imaginary part of (passing to continuous time)
\begin{equation}
\Pi(q,\omega) \equiv -i \int d{\bf r} \int_0^\infty dt\, e^{-i ({\bf q} \cdot {\bf r} + \omega t)}
f(r,t).
\end{equation}
For random walks $\Pi(q,\omega) \propto (\omega - i D q^2)^{-1}$ and
$\omega \Pi$ is a function of $q^2/\omega$ (diffusive scaling); such
scaling is violated at the semiclassical Hall transition, as seen
below.

The weight $2^{-H}$ in (\ref{ensemble}) is exactly that of the
$\theta^\prime$ model studied by Duplantier and
Saleur~\cite{duplantier}, which is in the same universality class as
the ordinary $\theta$-point.  The critical exponents $\gamma$ and
$\nu_\theta$, defined through
\begin{eqnarray}
\sum_{{\rm SAWs\ of\ length\ N}} {2^{-H}} &\sim& \mu^N N^{\gamma-1} \cr
{\sum_{{\rm SAWs\ of\ length\ N}} {R^2 2^{-H}} \over
\sum_{{\rm SAWs\ of\ length\ N}} {2^{-H}}}
&=& \langle R^2 \rangle_{\rm SAW} \sim N^{2 \nu_\theta},
\end{eqnarray}
take values $\gamma = \frac{6}{7}$, $\nu_\theta = \frac{4}{7}$ for this
class, and $\mu = 1$ on the hexagonal lattice~\cite{duplantier}.  Here
and in the following $\langle \rangle_{\rm SAW}$ denotes an average over
self-avoiding walks, while $\langle \rangle$ denotes an average over
disorder configurations in the random level surface problem.

The linear increase with time of the mean squared particle
displacement at $E_c$ follows from the values of polymer exponents
$\nu_\theta$ and $\gamma$: $\langle R^2(N) \rangle \propto N^{\gamma-1+2\nu_\theta}
= N.$  Higher moments of the particle distribution function
show nontrivial scaling laws:
\begin{equation}
\langle R^{2 n}(N) \rangle \propto N^{\gamma-1} N^{2 n\nu_\theta} = N^{(8
n-1)/7}, \quad n\geq 1.
\label{slaws}
\end{equation}
For random walks $\langle R^{2 n}(N) \rangle \propto N^{n}$.  The
higher moment laws (\ref{slaws}) reflect the non-intersection and memory
properties of random level surfaces, which lead to a non-Gaussian
distribution of $f(r,N)$.

The expression (\ref{ensemble}) for the particle distribution function
after $N$ steps includes both SAWs and SAPs.  Since the sum over SAPs
contains not just loops of length $N$ but of all shorter lengths, this
term contributes a large constant background which is difficult to
subtract numerically.  The most comprehensive numerics have been
performed on a reduced problem including only the average over
self-avoiding walks.  It follows from simple properties
of a polymer scaling function that Monte Carlo numerics suggesting
$\eta \approx 0$ for the reduced problem are not reaching the
asymptotic regime: the actual value is $\eta = \frac{1}{4}$ with a
logarithmic prefactor.  Then a similar argument gives $\eta =
\frac{1}{4}$ for the full problem (open and closed polymers), with no
logarithmic corrections.

The basic scaling law for polymers predicts that the
particle distribution function $f_0$ for the reduced problem has the
form
\begin{equation}
f_0({\bf r},t) = t^{-d \nu_\theta} H(r t^{-\nu_\theta})
\end{equation}
where $H$ is normalized as $\int H(x)\,d^d x = 1.$ $H$ is positive,
falls off rapidly as $x \rightarrow \infty$, and is smooth except for
the origin, where $H(x) \sim x^{(\gamma-1)/\nu_\theta} =
x^{-1/4}$.  This behavior at small $x$ is the simplest
example of a polymer contact exponent: a new critical exponent
describes the scaling as the two ends of the polymer approach each
other.  The value $(\gamma-1)/\nu_\theta$ can be derived from noting
that as the ends approach each other, a SAW becomes a SAP.  This
divergence at small $x$ is the source of the logarithmic corrections in
the reduced problem.

It is convenient to isolate the divergence in
$H(x)$:
\begin{equation}
H(x) = c x^{-1/4} e^{-x^2/2} + H_{\rm reg}(x),
\label{separate}
\end{equation}
where $c$ is a positive constant and $H_{\rm reg}$ is some smooth
function (no longer necessarily positive).  The cutoff on the singular
part is arbitrary; a Gaussian is chosen for simplicity.
The distribution function in momentum space is
\begin{eqnarray}
{\hat f}_0&({\bf q}&,t) = \int e^{-i {\bf q} \cdot {\bf r}}
f_0({\bf r},t)\, d^2 {\bf r}\cr
&=& 2 \pi \int_0^\infty J_0(q r) H(r t^{-4/7}) t^{-8/7}r\,dr
\equiv {\hat f}_0 (q t^{4/7}).
\end{eqnarray}

The contribution of the regular part $H_{\rm reg}$ to $f(q,\omega)$
is finite (possibly zero) as $\omega \rightarrow 0$:
\begin{eqnarray}
f_{\rm reg}(q,0) &\equiv&
2 \pi \int_0^\infty \int_0^\infty J_0 (u q t^{4/7}) H_{\rm reg}(u) u\,du\,dt \cr
&=& {7 \Gamma(7/8)^2 \sqrt{\sqrt{2}-1} \over 2 q^{7/4}}
\left(\int_0^\infty \frac{H_{\rm reg}(u)}{u^{3/4}}\,du\right).
\label{regpart}
\end{eqnarray}
The singular part of $H(x)$ gives a logarithmically divergent
contribution $C$ as $\omega \rightarrow 0$:
\begin{eqnarray}
C&=&2 \pi c \int_0^\infty \int_0^\infty e^{-i \omega t} J_0 (u q t^{4/7})
u^{-1/4} e^{-u^2/2} u\,du\,dt \cr
&=& 2 \pi c \Gamma(7/8) 2^{-1/8}
\int_0^\infty e^{-i \omega t} {}F(7/8,1,-q^2 t^{8/7}/2) \,
dt \cr
&=& {2 c \Gamma(7/8)^2 \sqrt{\sqrt{2}-1} \over 
q^{7/4}} \left(\log(q^{7/4} / \omega) + \ldots\right)
\label{singpart}
\end{eqnarray}
where the omitted terms are finite as $\omega \rightarrow 0$ and
$F$ is the confluent hypergeometric function.

From (\ref{singpart}) ${\rm Im\ }\Pi_0$ scales as $q^{-7/4} =
q^{-2+\eta}$ with $\eta = \frac{1}{4}$, with a logarithmically
divergent prefactor resulting from the polymer contact singularity
$H(x) \sim x^{-1/4}$.  The singularity is easily missed numerically
since it only starts to dominate the scaling function for small values
of $x$ in dimensionless units, so extremely long walks with $N \geq
10^4$ are required for its observation.  We have performed Monte Carlo
simulations of walks up to $N = 10^5$ to verify the predicted increase
of $H(x)$ at small $x$.

Now consider the average over all (closed and open) paths which gives
${\rm Im\ }\Pi$.  After a long time $t$, only a fraction
$t^{-\frac{1}{7}}$ of paths are open, and most particles move on
closed loops.  Na\"\i ve scaling predicts that $\eta = 2 x_1 =
\frac{1}{2}$, but the scaling function is shown below to vanish in the
limit $\omega \rightarrow 0$.  The next term is nonvanishing and gives
${\rm Im\ } \Pi \sim \omega^{1/7} q^{-7/4}$, so $\eta = 1/4$.  There
are no logarithmic corrections, unlike in the open-walk case, because
of the $t^{-\frac{1}{7}}$ damping of open walks.

As $t \rightarrow \infty$ the particle distribution
function goes to a nonzero limit
\begin{equation}
f(r,\infty) \equiv
\sum_{{\rm closed\ paths}\atop{\rm including\ 0\ and\ }r} 2^{-H} L^{-1},
\end{equation}
with $\sum_r f(r,\infty) = 1$ and $L$ is the path length.  The
background contributes a $\delta(\omega)$ part which is henceforth
ignored (difficulty in separating $\delta(\omega)$ is the reason why
Monte Carlo calculations are often performed on the reduced open-walk
problem).  Defining ${\tilde f}(r,N) \equiv f(r,N) - f(r,\infty)$,
some straightforward cancellations show that $\sum_N {\tilde f}(r,N) = 0$ for
all $r$, so for all $q$, in the continuum limit $\int_0^\infty {\tilde
f}(q,t)\, dt = 0$ and ${\rm Im\ }\Pi \rightarrow 0$ as $\omega
\rightarrow 0$.

For nonzero $\omega \ll q^{7/4}$ the contribution of paths shorter
than $\omega^{-1}$ is nearly zero from the same cancellations; long
paths determine
\begin{equation}
{\rm Im\ }\Pi \sim \int_{\omega^{-1}}^\infty t^{-1/7}
\cos(\omega t-1) g(q t^{4/7})\,dt
\sim \frac{\omega^{1/7}}{q^{7/4}}.
\end{equation}

The upshot of the above results is that for neither the open-walk case
nor the full problem is the scaling trivial: logarithmic corrections
appear in the open-walk case, and the scaling function in the full
problem vanishes for small argument, leading to the emergence of a new
power-law and $\eta = \frac{1}{4}$ instead of $\eta = 2 x_1 =
\frac{1}{2}$.  A similar discrepancy between $\eta$ and the value of
$2 x_1$ obtained from finite-size scaling for the integer Hall transition
was noted in the original paper~\cite{chalker}.

An interesting question is whether the spin quantum Hall
transition (SQHT), which has the same $\sigma$ and $\nu$ as the
semiclassical transition (Table I) assuming a simple relation between
Landauer and Einstein conductances~\cite{gurarie}, also has the same
$\eta$.  Some sums over paths in the SQHT network model are
exactly equal to sums over percolation
hulls~\cite{gruzberg,beamond}, but individual paths are not directly
related to individual hulls, so it is not clear that $\eta$ need
be the same.

\begin{table}
\begin{tabular}{|c|c|c|c|}
\makebox[5mm]{\ }&\makebox[5mm]{$\nu$}
&{$\sigma$}&\makebox[5mm]{$\eta$}\cr
\cline{1-4}&&&
\\ CQHT&$\frac{4}{3}$~\cite{trugman}&
$\frac{\sqrt{3}}{4} \approx 0.433$~\cite{cardy}&$\frac{1}{4}$\\
IQHT&$2.35 \pm 0.05$~\cite{huckestein}&$0.5 \pm 0.1$~\cite{huo}&
$0.38 \pm 0.04$~\cite{chalker}\\
SQHT&$\frac{4}{3}$~\cite{gruzberg}&$\frac{\sqrt{3}}{4}$~\cite{cardy}&
?\\
\end{tabular}
\caption{Comparison of exact critical properties of semiclassical Hall
transition with the ordinary integer transition (IQHT) and spin transition
(SQHT).  Critical conductivity is in units of $\frac{e^2}{h}$ per spin.}
\label{tab1}
\end{table}

Several authors~\cite{shimshoni,kapitulnik} have proposed that the
semiclassical limit discussed above may be relevant to the many
experimental samples which fail to show simple $\nu \approx 2.3$
scaling down to low temperatures.  The
exponent $\nu$ is traditionally obtained by measuring two different
combinations of $\nu$ and $z$, where the dynamical exponent $z$ is
typically equal to 1 within experimental error.  The scaling of the
width of the transition region with temperature measures the product
$\nu z$: experimental results on such scaling show variously scaling
down to low temperature with $\nu z \approx 2.3$~\cite{wei}, scaling
down to low temperature with $\nu z \approx 1.5$~\cite{hohls}, and a
breakdown of scaling at low temperature with high-temperature scaling
$\nu z \approx 1.5$~\cite{shahar}.

An alternate way to determine the localization length and hence $\nu$
directly, without measuring $z$, is via variable-range hopping in the
localized regime far from the transition.  Recent
experiments~\cite{hohls} found for three samples one scaling law for
the plateau width, corresponding to $\nu \approx 1.5$ if $z=1$, and
another corresponding to $\nu \approx 2.3$ for the localization length
obtained from variable-range hopping at finite temperature.  Other samples
had plateau width $\nu$ values ranging from 1.3 to 2.2.  This
surprising appearance of different power laws in the same samples
suggests either a nontrivial value for $z \neq 1$ or that the two
measurements are probing different physics.  The remainder of this
paper discusses an interpretation of these experiments based on
quantum-mechanical tunneling between semiclassical states and a loss
of phase coherence near the critical energy.  Note that the single-particle motion
on percolation hulls discussed below is distinct from percolation of
macroscopic quantum Hall ``puddles''~\cite{shimshoni}.

A possible explanation for the existence of two scaling laws in the
same samples is that the plateau width measurements, taken at high
current so that each electron state is only briefly occupied, do not
see the effects of the relatively slow quantum tunneling processes,
which cause $\nu \approx 2.3$ for the true zero-temperature,
infinite-time localization length.  The effective localization length
on short time scales should then be described by the classical
percolation exponent $\nu = \frac{4}{3}$.  In~\cite{hohls} the plateau
width measurements are taken at conductivity $\sigma$ of order
$10^{-4} e^2/h$, while the variable-range-hopping localization length
is taken from data over the range $10^{-13} \leq \sigma h/e^2 \leq
10^{-5}$.  This argument predicts is that the localization length
extracted from plateau width scaling should be shorter than that
from variable-range hopping at low current.

An area of current interest is how dissipation via
coupling to low-energy excitations, such as weakly
localized electrons, can modify quantum phase
transitions~\cite{kapitulnik}.  Such low-energy excitations of unknown origin seem
experimentally to generate a finite dephasing length down to the lowest
measured temperatures in some samples.
As the chemical potential nears the critical energy, power-law scaling of the
localization length requires phase coherence on increasing length and
time scales.  The physics of the quantum Hall fixed point depends on
both tunneling and phase coherence, which keeps the states from being
truly extended except at the critical energy.  Once the localization
length is larger than the dephasing length, the transport should be
diffusive (finite $\sigma_{xx}$), as seen in some samples in a finite
range around the critical energy.

Another explanation for classical percolation scaling is that a smooth
disorder potential (disorder correlation length larger than the
magnetic length) shows a larger crossover region where classical
percolation applies than a sharp potential, because tunneling is
reduced.  However, this would not explain the breakdown in scaling of
the plateau width at low temperature observed in many samples, and
would require samples from similar growth runs to have very
different impurity distributions.  A possible direct test of whether changes
in dephasing length are indeed correlated with the sample-dependent
behavior would be the addition of a controllable coupling to dissipation, as
done for low $B$ in~\cite{rimberg}.

Our main conclusion is that the semiclassical limit of the quantum Hall transition has nontrivial mobility edge scaling which can be found exactly using results from the theory of polymers or percolation.  This limit can be experimentally observed in samples with dissipation or smooth potential fluctuations and is closely related to the spin quantum Hall transition.

{\it Note added:} After this paper was submitted, a preprint~\cite{eversnew} appeared
which examines the anomalous scaling exponent at the spin quantum Hall transition and also finds $\eta=\frac{1}{4}$.

The author thanks I. Affleck, A. Green, I. Gruzberg, V. Gurarie, and
A. Vishwanath for useful conversations.

\bibliography{ranlev4}
\end{document}